\documentclass[11pt]{article}
\usepackage[cp866]{inputenc}

\newcommand{\const}{{\rm\, const}}

\makeatletter

\renewcommand{\@oddhead}{\hfil Alexander Shatskiy }

\input epsf

\begin{document}

\begin{center}
\large {\bf Influence of Rotation on the Amount of Phantom Matter
around wormholes}\\
Alexander Shatskiy \footnote{Astro Space Center, Lebedev Physical
Institute, Moscow, Russia; shatskiy@asc.rssi.ru }
\end{center}

PACS numbers : 95.30.Sf
$$ $$
{\bf \Large Abstract}\\
Static (spherically symmetrical) and stationary solutions for
wormholes are considered. The visibility horizon, which
characterizes the differences between black holes and passing
wormholes, is determined in an invariant way. It is shown that the
rotation of wormholes does not affect the amount of phantom matter
that surrounds them.

\section{INTRODUCTION}
\label{s1}

In recent work in relativistic astrophysics, interest in solutions
with wormholes (WHs) has increased, due, in particular, to the
future construction of high accuracy radio interferometers that
will be able to distinguish WHs from other objects (such as black
holes).

A fundamental and characteristic property of a WH is its throat,
through which physical bodies can pass. Spacetime is strongly
curved near the throat. This curvature reaches values
corresponding to the horizon of a black hole with the same mass.

A necessary condition distinguishing a black hole from a WH is the
absence of an event horizon for the latter. Therefore, in Section
\ref{s4}, we find an invariant criterion for the existence of an
event horizon in an axially symmetrical system.

These properties of WHs are maintained by the presence of exotic
matter surrounding the throat. Due to this exotic matter, the
radial pressure near the WH throat is less than the negative of
the energy density: ${p_\parallel < -\varepsilon}$ (a super-rigid
equation of state). This matter is called phantom matter. We show
in Sections \ref{s4}-\ref{s7} that rotation of the WH does not
influence the required amount of surrounding phantom matter.

\section{SPHERICALLY SYMMETRICAL CASE}
\label{s2}

The solution for WHs in the spherically symmetrical case was
studied in detail, for example, in \cite{2}, where it was shown
that the spherically symmetrical solution for a WH makes a smooth
transition to the solution at the horizon\footnote{The definition
of the horizon for the more general case will be given in Section
\ref{s4}.} and the WH becomes unpassable.

The case of the small difference of a WH from an electrically (or
magnetically) charged Reisner-Nordstrem black hole may be of
practical interest~\cite{3}.

Let us consider matter with a linear equation of state
${(p=\const\cdot\varepsilon)}$, when the coefficients of the
energy density are constant and equal in magnitude to
\begin{equation}
w \equiv 1+\delta =-p_\parallel /\varepsilon=p_\perp /\varepsilon
\, ,\quad x=r/r_h \, ,\quad \xi
(x)=\varepsilon(r)/\varepsilon(r_h) \, ,\quad y(x)=x\cdot
[1+1/g_{rr}(x)]\, . \label{6-1}\end{equation} Here, $r_h$ is the
radius of the horizon for a black hole with the same total mass as
the WH. We now write the metric for the WH, which has, in the
general case, the form
\begin{equation}
ds^2= \exp(2\phi)\cdot [dt^2] - (1-y(x)/x)^{-1}\cdot [dr^2] -
r^2\cdot \left( [d\theta^2] + \sin^2\theta\, [d\varphi^2]\right)
\, . \label{6-2}\end{equation} The equations for this metric can
be written in the convenient form~\cite{2}:
\begin{equation}
\begin{array}{ccc}
y(x) = 2-\int\limits_x^\infty \xi\, x^2\, dx / w \\
{1\over\xi}{d\xi\over dx} = -4/x + {\delta\over 2w}\cdot (\xi x
-y/x^2)/
(1-y/x)\\
\exp [\phi (x)] = const / (\xi x^4)^{w/\delta}
\end{array}
\label{6-3}\end{equation} When ${\delta=0}$, this solution makes a
transition to the solution of Reisner-Nordstrem solution
containing the horizon: ${y_{_{[\delta=0]}}(x)=2-1/x\, ,\quad
\xi_{_{[\delta=0]}}(x)=1/x^4\, , \quad
\exp[\phi_{_{[\delta=0]}}(x)]=1-1/x .}$ We obtain for the solution
(\ref{6-3}) an analytical expression representing a first
approximation in the small correction $\delta$. We denote
${z\equiv(1-1/x)}$. In a linear approximation in $\delta$,
Eq.~(\ref{6-3}) for $\xi$ can be re-written in the form
\begin{equation}
{\partial \ln (\xi\, x^4) \over \partial x} = -\delta \cdot
{\partial \ln (z) \over \partial x} \label{6-4}\end{equation}
Taking into account the requirement that $\xi$ concide with
${\xi_{_{[\delta=0]}}}$ at infinity, we obtain using this relation
\begin{equation}
\xi\, x^4 = 1/z^{\delta} \, , \quad y(x) = 1+ z^{1-\delta} \, ,
\quad \exp (\phi) = z^{1+\delta} \, . \label{6-5}\end{equation}
The radius of the throat ${r_0=r_h\cdot x_0}$ is determined by the
transcendental equation ${y(x_0)=x_0}$, from which we can obtain
\begin{equation}
\delta = -\ln (x_0)/\ln (1-1/x_0) \, . \label{6-6}\end{equation}
The parameter $\delta$ is determined by the equation of state of
the matter in the WH, including quantum corrections: ${\delta
\equiv w -1 >0}$. For example, when ${x_0=1.001}$, ${\delta\approx
1.4\cdot 10^{-4}}$.

\section{ROTATING WORMHOLES}
\label{s4}

In the simplest case, the metric tensor for a rotating WH can be
obtained by adding a non-diagonal term to the spherically
symmetrical WH metric. It is convenient to replace the $dr^2$ term
in (\ref{6-2}) by introducing a new variable $l$ that is equal to
zero at the throat:
\begin{equation}
dl^2 \equiv (1-y(x)/x)^{-1}\, dr^2 \label{1-0}\end{equation} The
metric then takes the form
\begin{equation}
ds^2= g_{tt}(l) \cdot [dt^2] - [dl^2] - r^2 (l) \cdot \left(
[d\theta^2] + \sin^2\theta\, [d\varphi^2]\right)  + 2
f(l)\cdot\sin^2\theta \cdot [dt\cdot d\varphi] \, .
\label{1-1}\end{equation} The WH metric can be written in the form
(\ref{1-1}) only in the case of slow rotation. Otherwise, a
$\theta$ dependence will appear in the metric coefficients, apart
from the fact that this dependence also appears in $g_{t\varphi}$
and $g_{\varphi\varphi}$ (see~\cite{6},\cite{7}).

We introduce the required notation for the cylindrical coordinate
$\rho$ and the modulus of the metric tensor $g$:
\begin{equation}
\rho\equiv r(l) \, \sin\theta \, , \quad g\equiv g_{tt}\, r^2\,
\rho^2 + r^2\, g^2_{t\varphi}\, . \label{1-2}\end{equation} A
prime will denote a derivative with respect to $l$.

A necessary condition for the existence of a WH is the absence of
an event horizon. The physical meaning of the event horizon is
defined in \cite{5}.

The mathematical definition of the event horizon is the zero
hypersurface having the property that it transmits the world lines
of moving particles in only one direction~\cite{4}. Thus, a length
element $ds$ should be equal to zero in the direction normal to
this hypersurface. Precisely this property implies that world
lines of particles or light rays (directed into the future) can
cross this hypersurface in only one direction.

We now introduce the concept of the invariant velocity $V$ , which
coincides with the usual three-dimensional velocity for moving
particles in the non-relativistic case, and is equal to unity at
the event horizon, independent of the chosen reference frame.

Thus, one definition for $V$ that satisfies the above properties
is the given by the expression
\begin{equation}
V^2\equiv 1-{ds^2\over dt^2}=1-{1\over \left( U^t \right)^2}\, ,
\label{1-3}\end{equation} where $U^i$ are the components of the
4-velocity of the particle. Note that ${V^2\to v^2}$ in the
non-relativistic limit, where $v$ is the usual non-relativistic
velocity.

To express the invariant velocity in terms of the metric
components, we use the fact that there exist four integrals of
motion for particles in an axially symmetrical field~\cite{3}. Two
of these integrals (for an uncharged particle) are $U_t$ and
$U_\varphi$.

${U_t = E_0/m}$  expresses the conservation of energy and
${U_\varphi =L/m}$ the conservation of angular momentum for the
particle.

Hence, (\ref{1-3}) can be re-written in the form
\begin{equation}
V^2 = 1-\left( {m\over E_0}\right)^2 \cdot {1\over \left( g^{tt} +
g^{t\varphi}\, L/E_0 \right)^2 }\, . \label{1-4}\end{equation}
Note that this definition of the event horizon is valid for any
axially symmetrical metric (including the Newman-Kerr
metric~\cite{3}).

In the case of a wormhole (see Table~1 in Appendix~\ref{app}), the
horizon is defined as the geometric locus of points for which
${g_{tt}+(f\,\sin\theta /r)^2=0}$ or ${g=0}$. This surface is egg
shaped\footnote{The coordinates for this surface can be reduced to
a spherical form via an appropriate transform; in this case, the
ergosphere will have the form of a flattened ellipsoid.} and
elongated along the axis of rotation.

The spherical surface ${g_{tt}(l)=0}$  lies outside the horizon,
touching it at the poles, and has the meaning of the outer
boundary of the ergosphere.

It follows from the above that the WH throat is located outside
the ergosphere, and the quantity ${1/{r'}^{2}}$ is simultaneously
for the metric (\ref{1-1}) the component ${-g_{rr}\to -\infty}$
for ${l\to 0}$. Then, particles falling through the throat will
not reach the horizon (which should be absent for a WH).

\section{THE EINSTEIN EQUATION}
\label{s6}

In the general case, it is convenient to write the Einstein
equation in the form
\begin{equation}
R_{ik} = 8\pi \left( T_{ik} - {1\over 2}g_{ik}\, T \right)  \, .
\label{3-1}\end{equation} We use $\sigma^i$ to denote the zero
4-vector of a photon with zero angular momentum relative to the
center of the system. Since the angular momentum is determined by
the covariant component of the 4-velocity, ${\sigma_\varphi = 0}$
by definition. We can then present $\sigma$ in the form
\begin{equation}
\sigma_i = F_0\cdot \{ 1/\sqrt{g^{tt}},\, 1,\, 0,\, 0 \}\, ;\quad
\sigma^i = F_0\cdot \{ \sqrt{g^{tt}},\, -1,\, 0,\,
g^{t\varphi}/\sqrt{g^{tt}} \}\, ;\quad \sigma_i\, \sigma^i =0  \,
, \label{3-2}\end{equation} where $F_0$ is any function. It is
convenient to choose this function to be ${F_0=\sqrt{g_{tt}
g^{tt}}}$. In this case, ${\sigma_i u^i =1}$, where $u^i$ is the
4-velocity of the matter in the comoving reference frame (see
Table~\ref{T2} in Appendix~\ref{app}).

Performing a scalar multiplication of both sides of (\ref{3-1}) by
the 4-vector $\sigma^i$ twice yields
\begin{equation}
R_{ik}\, \sigma^i\, \sigma^k = 8\pi T_{ik}\, \sigma^i\, \sigma^k
\equiv 8\pi \Pi \, . \label{3-3}\end{equation} The scalar $\Pi$ is
directly proportional to the energy density measured by an
observer in a reference frame tied to the photon. It stands to
reason that this reference frame is meaningless, or rather has
meaning only in the asymptotic limit for a system whose velocity
approaches the velocity of light.

In the special case of the energy-momentum tensor $T_{ik}$ of an
ideal fluid (see Table~\ref{T2} in Appendix~\ref{app}), this
scalar is equal to ${\Pi = \varepsilon + p}$.

Its physical meaning for ordinary matter means that $\Pi$ cannot
be negative. This requirement is called the zero-energy condition
(NEC), and matter that violates this condition is called phantom,
or exotic, matter. To test the violation of the zero-energy
condition for a rotating wormhole, we must calculate the necessary
components of the Ricci tensor.

\section{THE CURVATURE TENSOR}
\label{s7}

The Ricci tensor has the form
$$
R_{ik} = \partial_n (\Gamma^n_{ik}) - \partial_k (\Gamma^n_{in}) +
\Gamma^n_{ik}\cdot\Gamma^m_{nm} - \Gamma^m_{in}\cdot\Gamma^n_{km}
$$
The required Cristoffel symbols are
$$
\Gamma^i_{kn}={1\over 2}g^{im}\, (\partial_n g_{mk} + \partial_k
g_{mn} - \partial_m g_{kn}),
$$
and the inverse metric tensor $g_{ik}$ was calculated and
presented in Tables~\ref{T1} and \ref{T3} in Appendix~\ref{app}.

Using these results, we obtain
\begin{eqnarray}
R_{tt}= {{g''}_{tt}\over 2} - {\rho^2 (ff' \, g'_{tt} - {f'}^2\,
g_{tt})\over 2g} + {f^2\, g_{tt}\, \sin^2 (2\theta)\over 2g} +
{\rho^2\, g'_{tt}\, (g_{tt}\, (r^2)' - g'_{tt}\, r^2)\over 4g} +
{g'_{tt}\, r'\over 2r} \, ,
\label{2-1}\\
R_{ll}= -\left[ {\rho^2 ((r^2\, g_{tt})' + 2 ff' \, \sin^2\theta
)\over 2g}\right]'
-\left[ {\rho^2 (r^2\, g'_{tt} + ff'\, \sin^2\theta )\over 2g} \right]^2 -
\nonumber \\
- \left[ {\rho^2 (ff' \, \sin^2\theta + 2 g_{tt} r r' )\over 2g}
\right]^2 - {r^2\, \rho^6 (f/r^2)'  (f\, g'_{tt} - f'\,
g_{tt})\over 2g^2} - {r''\over r} \, ,
\label{2-2}\\
R_{t\varphi}= {f''\, r^2\, \sin^2 \theta + 2 f \, \cos (2\theta) +
f'\, r\, r'\, \sin^2 \theta \over 2 r^2} -
\nonumber \\
- { (f'\, g'_{tt}\, r^2 + f\, {f'}^2\,)\, r^2\, \sin^4 \theta +
f\, g_{tt} \, r^2 \, \sin^2 (2\theta) +(2\, f'\, g_{tt} - 4\, f\,
g'_{tt})\, r^3\, r'\, \sin^4 \theta \over 4 g} \, ,
\label{2-5}\\
R_{\varphi\varphi}= {2\rho^4 {r'}^2 g_{tt} + 0.5 r^4 g_{tt} \sin^2
(2\theta)
- rr' \rho^4 g'_{tt} - \rho^6 f' (f/r^2)' \over 2 g} -
\nonumber \\
- {r''\over r} \rho^2 - 2 {r'}^2 \sin^2\theta - \cos (2\theta)\, .
\label{2-6}\end{eqnarray}

\section{PHANTOM MATTER NEAR A WORMHOLE}
\label{s8}

In a spherically symmetrical WH, phantom matter appears near the
throat~\cite{1}. Using (\ref{3-2}-\ref{2-6}), we can find the
value of the scalar at the throat $(\Pi_0)$ for a rotating WH,
taking into account that ${r'=0}$ at the throat, and any
derivative with respect to $l$ can be replaced as follows:
${{\partial\over \partial l} \to r' {\partial\over \partial r}}$.
Therefore, at the throat, we must include only terms without
derivatives and with second derivatives with respect to $l$.

We thus obtain at the throat (in linear approximation in respect
to $f^2$):
\begin{equation}
\Pi_0 = F_0^2 \left[ -{2r''\over r} + {f^2 (3\cos^2\theta -1)
\over g_{tt} r^4} \right] \label{4-1}\end{equation} Using
reasoning analogous to the spherically symmetrical case~\cite{1},
we can convince ourselves that there will also always be a region
with phantom matter for a rotating WH. The integral of $\Pi_0$
over the all angle ${d\Omega =\sin\theta \,
d\theta|_0^{\pi}d\varphi|_{-\pi}^{\pi}}$ yields the same value for
the quantity of phantom matter near the throat as in the absence
of rotation. It is possible that this result is due to the
smallness of the rotation, and will not be valid for more extreme
cases.

\section{CONCLUSION and ACKNOWLEDGMENTS}
\label{s9}

In a linear approximation in the angular velocity, the rotation of
a WH does not influence the amount of phantom matter surrounding
the WH.

\bigskip
\hrule height 1pt
\bigskip

I would like to thank N.S. Kardashev, I.D. Novikov and R.F.
Polishchuk for fruitful discussions of this work and for valuable
comments. This work was supported by the Russian Foundation for
Basic Research (project codes: 05-02-17377, 05-02-16987-a,
05-02-17257-a, NSh-1653.2003.2).

\section{Appendix (tables)}
\label{app}

\begin{center}
\large

{\bf Metric tensor:}
\begin{equation}
\label{T1}
\begin{tabular}{|c||c|c|c|c|||c|||c||c|c|c|c|}
\hline

$g_{ik}$ & $t$ & $l$ & $\theta$ & $\varphi$ & \quad &
$g^{ik}$ & $t$ & $l$ & $\theta$ & $\varphi$\\
\hline\hline $t$         & $g_{tt}$ & 0 & 0 & $f\,\sin^2 \theta$ &
\quad &
$t$ & $r^2\, \rho^2 / g$ & $0$ & $0$ & $f \, \rho^2 / g$  \\
\hline $l$         &  $0$  & $-1$ & $0$ & $0$ &\quad &
$l$   & $0$  & $-1$  & $0$ & $0$  \\
\hline $\theta$ &  $0$ & $0$ &  $-r^2$ & $0$ & \quad &
$\theta$ & $0$ & $0$ & $-1/r^2$ & $0$  \\
\hline $\varphi$ & $f\,\sin^2 \theta$ & $0$ & $0$ & $-\rho^2$ &
\quad &
$\varphi$ & $f \, \rho^2 / g$ & $0$ & $0$ & $-r^2 \, g_{tt} / g$ \\

\hline
\end{tabular}
\end{equation}

{\bf Energy-momentum tensor for an ideal fluid in a frame co-moving with the matter:}\\
${T_{ik}=(\varepsilon + p) u_i u_k - p g_{ik}\, ;\,\, u^i = \{
1/\sqrt{g_{tt}}, 0, 0, 0 \}\, ; \,\, u_i = \{ \sqrt{g_{tt}}, 0, 0,
g_{t\varphi}/\sqrt{g_{tt}} \}\, .}$
\begin{equation}
\label{T2}
\begin{tabular}{|c||c|c|c|c|}
\hline

$T_{ik}-{1\over 2}g_{ik}T$ & $t$ & $l$ & $\theta$ & $\varphi$\\
\hline\hline $t$  & ${1\over 2}(\varepsilon +3p) g_{tt}$ &
$0$ & $0$ & ${1\over 2}(\varepsilon +3p) g_{t\varphi}$ \\
\hline
$l$ & $0$ & ${1\over 2}(\varepsilon - p)$ & $0$ & $0$   \\
\hline
$\theta$ & $0$ & $0$ & ${1\over 2} r^2 (\varepsilon - p)$ & $0$ \\
\hline $\varphi$ & ${1\over 2}(\varepsilon +3p) g_{t\varphi}$ &
$0$ & $0$ &
$(\varepsilon + p) g^2_{t\varphi}/g_{tt} + {1\over 2} \rho^2 (\varepsilon - p)$\\

\hline
\end{tabular}
\end{equation}

{\bf Cristoffel symbols:}
\begin{equation}
\label{T3}
\begin{tabular}{|c||c|c|c|c|}
\hline

$\Gamma^t_{ik}$ & $t$ & $l$ & $\theta$ & $\varphi$\\
\hline\hline $t$  & $0$ & ${\rho^2 (r^2\, g'_{tt} + ff'
\sin^2\theta )\over 2g}$ &
${f^2\,\rho^2\,\sin(2\theta) \over 2g}$ & $0$ \\
\hline $l$ & ${\rho^2 (r^2\, g'_{tt} + ff' \sin^2\theta )\over
2g}$ & $0$ & $0$ &
${r^2\,\rho^4\, (f/r^2)' \over 2g}$   \\
\hline
$\theta$ & ${f^2\,\rho^2\,\sin(2\theta) \over 2g}$ & $0$ & $0$ & $0$ \\
\hline
$\varphi$ & $0$ & ${r^2\,\rho^4\, (f/r^2)' \over 2g}$ & $0$ & $0$ \\

\hline \hline \hline \hline

$\Gamma^l_{ik}$ & $t$ & $l$ & $\theta$ & $\varphi$\\
\hline\hline
$t$   & ${1\over 2} g'_{tt}$ & $0$ & $0$ & ${1\over 2} f' \sin^2\theta$ \\
\hline
$l$  & $0$ & $0$ & $0$ & $0$ \\
\hline
$\theta$ & $0$ & $0$ & $-r r'$ & $0$ \\
\hline
$\varphi$ & ${1\over 2} f' \sin^2\theta$ & $0$ & $0$ & $-r r' \sin^2\theta$ \\

\hline \hline \hline \hline

$\Gamma^\theta_{ik}$ & $t$ & $l$ & $\theta$ & $\varphi$\\
\hline\hline
$t$  & $0$ & $0$ & $0$ & ${f \sin (2\theta) \over 2 r^2}$ \\
\hline
$l$ & $0$ & $0$ & $r'/r$ & $0$ \\
\hline
$\theta$ & $0$ & $r'/r$ & $0$ &   \\
\hline
$\varphi$ & ${f \sin (2\theta) \over 2 r^2}$ & $0$ & $0$ & $-{1\over 2}\sin (2\theta)$ \\

\hline \hline \hline \hline

$\Gamma^\varphi_{ik}$ & $t$ & $l$ & $\theta$ & $\varphi$\\
\hline\hline $t$ & $0$ & ${\rho^2 (f g'_{tt} - f' g_{tt})\over
2g}$ &
${f r^2 g_{tt} \sin (2\theta) \over -2g}$ & $0$ \\
\hline $l$  & ${\rho^2 (f g'_{tt} - f' g_{tt})\over 2g}$  & $0$ &
$0$ &
${\rho^2 (ff' \sin^2\theta +2g_{tt} rr') \over 2g}$ \\
\hline $\theta$ & ${f r^2 g_{tt} \sin (2\theta) \over -2g}$ & $0$
& $0$ &
${r^2 \sin (2\theta) (g_{tt} r^2 + f^2 \sin^2\theta) \over 2g}$ \\
\hline $\varphi$ & $0$ & ${\rho^2 (ff' \sin^2\theta +2g_{tt} rr')
\over 2g}$ &
${r^2 \sin (2\theta) (g_{tt} r^2 + f^2 \sin^2\theta) \over 2g}$ & $0$ \\

\hline

\end{tabular}
\end{equation}

\end{center}


Translated by D. Gabuzda


\begin{thebibliography}{99}

\bibitem{2}
A.A. Shatskii, Astron. Zh. {\bf 81}, p.579 (2004) [Astron. Rep.
{\bf 48}, p.525 (2004)].

\bibitem{3}
C. W. Misner, K. S. Thorne, and J. A. Wheeler, {\it Gravitation},
Vol.3, (Freeman, San Francisco, 1973; Ainshtain, Moscow, 1997).

\bibitem{5}
A.A. Shatskii, Zh. Eksp. Teor. Fiz. {\bf 116}, 353 (1999) [JETP
{\bf 89}, 189 (1999)].

\bibitem{4}
L.D. Landau and E.M. Lifshitz, {\it The Classical Theory of
Fields} (Nauka, Moscow, 1995; Pergamon, Oxford, 1975), Vol. 2.

\bibitem{1}
Matt Visser, {\it Lorentzian Wormholes From Einstein to Hawking}
(United Book, Baltimore, 1996).

\bibitem{6}
Ya. B. Zel'dovich and I. D. Novikov, {\it The Theory of
Gravitation and Stellar Evolution} (Nauka,Moscow, 1971) [in
Russian].

\bibitem{7}
V.P. Frolov, I.D. Novikov, {\it Black Hole Physsics. Basic
Concepts and New Developments}, Kluver AP (1998).

\end{thebibliography}
\end{document}